\documentclass[12pt]{iopart}

\newcommand{\ket}[1]{\left\vert#1\right\rangle}
\newcommand{\bra}[1]{\left\langle#1\right\vert}

\newcommand{\eq}{Eq.~}
\newcommand{\eqs}{Eqs.~}

\newcommand{\cf} {cf.~}
\newcommand{\ug} {\!=\!}

\newcommand{\piu} {\!+\!}
\newcommand{\meno} {\!-\!}

\newcommand{\rref} {Ref.~}
\newcommand{\rrefs} {Refs.~}
\usepackage{bm,color}

\begin{document}

\title{A quantum non-Markovian collision model: incoherent swap case}

\author{F. Ciccarello$^{1,2}$ and V. Giovannetti$^1$ }

\address{$^1$ NEST, Scuola Normale Superiore and Istituto Nanoscienze-CNR, Piazza dei Cavalieri 7, I-56126 Pisa, Italy}
\address{$^2$ NEST, Istituto Nanoscienze-CNR and Dipartimento di Fisica e Chimica, Universit$\grave{a}$  degli Studi di Palermo, via Archirafi 36, I-90123 Palermo, Italy}



\ead{francesco.ciccarello@unipa.it}
\begin{abstract}
We have recently presented a collision-model-based framework to approach non-Markovian quantum dynamics [Ciccarello F Palma G M and Giovannetti V 2013 {\it Phys.~Rev.~A} {\bf 87} 040103(R)]. As a distinctive feature, memory is introduced in a dynamical way by adding extra inter-ancillary collisions to a standard (memoryless) collision model. Here, we focus on the case where such intra-bath collisions are described by incoherent partial swap operations. After briefly reviewing the model, we show how to include temperature as an additional parameter by relaxing the assumption that each bath ancilla is initially in a pure state. We also calculate explicitly the dynamical map entailed by the master equation in the paradigmatic instance of a Jaynes-Cummings system-ancilla coupling.
\end{abstract}

\maketitle

\section{Introduction}

The theoretical settling of open system dynamics ranks among the hottest problem in modern quantum mechanics \cite{petruccione,weiss,huelga}. By definition, an open system is one in contact with an external environment and thus the dynamical map describing its time evolution is non-unitary. On a rather general basis, a physically grounded dynamical map is required to be completely positive and trace preserving (CPT). In the case that the environment is memoryless, a Markovian dynamics occurs which, demonstrably, is always governed by a so called Lindblad-type master equation (ME) \cite{petruccione,weiss,huelga}. Any Lindblad ME in turn corresponds to an unconditionally CPT dynamical. In the general case, though, environments are intrinsically {\it non-Markovian} (NM) and in several known scenarios a Markovian-based approach is fully inadequate \cite{nm-actual}. Unlike the Markovian case, however, a general systematic framework to tackle NM processes has not been developed to date \cite{breuer}. Rather, a number of variegated approaches have been proposed. Usually, the major drawback of these is that the corresponding MEs can violate the CPT condition in some regimes \cite{sabrina,laura,kossa}. Typically, this stems from the phenomenological assumptions and approximations that one somehow needs to make in order to derive a ME \cite{vacchini1}. A further recurrent drawback is that some MEs, such as those in \rrefs\cite{barnett, lidar}, focus on a regime which is too weak a perturbation of pure Markovianity.  As a result, their capability to capture genuinely NM features can be severely limited \cite{laura-breuer}.

Recently, we have proposed an innovative collision-model-based approach to tackle non-Markovian dynamics \cite{lavoro}. Collision models (CMs) were first envisaged by Boltzmann as an interesting way to describe the irreversible dynamics of a system in contact with a large environment, or bath, in terms of successive ``collisions" with its small subparts (ancillas). In more recent years, mostly, they were successfully applied to study quantum open systems \cite{rau,alicki,scarani, buzek,vittorio,indivisible}. In a standard CM, the bath is modeled as a large collection of non-interacting identical ancillas. $S$ (the system under study)  ``collides" with each of these one at a time. It can be shown that such a process gives rise to an irreversible dynamics for $S$
exactly described by a Lindblad-type ME {\cite{buzek,vittorio}}. This is because $S$ is not allowed to interact more than once with a given ancilla. The bath thus cannot keep track of the system's past history. 
A major reason why CMs are attractive is that they are suited to work out exact MEs basically without any approximations (hence ruling out the possibility to violate the CPT condition). Only the passage to the continuous limit is needed \cite{lavoro,buzek,vittorio,indivisible}. This is in contrast to standard microscopic system-reservoir models \cite{petruccione}, where even to derive Markovian MEs drastic assumptions (such as the Born-Markov approximation) are in fact unavoidable. Very recently, Rybar {\it et al.} introduced a NM CM able to simulate any indivisible single-qubit channel \cite{indivisible}. In their framework, memory is introduced by simply allowing for correlated initial bath states (aside from this, the model is identical to memoryless CMs).

In our model \cite{lavoro}, instead, the initial reservoir state is fully uncorrelated (just like in a standard Markovian CM) but ancillas can undergo pairwise collisions. This enables transmission of quantum information across the bath in a way that the initial state of a given ancilla before colliding with the system is affected by the $S$ past history. Moreover, this mechanism endows the reservoir with memory in a {\it dynamical} way: the ability to remember is associated with physical parameters entering the system-bath interaction, which is what one would intuitively expect.
In \rref\cite{lavoro}, we have proposed to describe each inter-ancillary collision as a partial swapping operation. This is indeed a natural choice to account for intra-bath information transfer. Moreover, it is defined in terms of the swap unitary operator, which allows to considerably simplify calculations. A possible choice is what we call an {\it incoherent swap}, i.e, a map whose Kraus operators are proportional to the identity and swap operators, respectively \cite{lavoro}. Using this, which is non-unitary but CPT, we have demonstrated that one can work out an exact ME interpolating between a fully Markovian regime and a strongly NM one. Importantly, such ME entails a dynamics which is unconditionally CPT \cite{lavoro}. Furthermore, it does not depend either on the system-ancilla coupling form or the dimensionality of the particles.

In this short paper, we consider the aforementioned incoherent swap CM and present some related developments with a twofold goal. First, our theory in \rref\cite{lavoro} was restricted to the case that the initial bath state is pure. Yet, any open system model is expected to feature temperature among its parameters. This is usually taken into account by considering the reservoir initially in a thermal, thus in general mixed, state. In this respect, we show that assuming an initial pure state is not restrictive provided that one redefines the ancillas as initially entangled bipartite systems where only one of the two can interact with $S$ {(an approach that in the theory of quantum channels~\cite{REV} corresponds to the Stinespring dialation mechanism)}. Our second aim is to illustrate the explicit form of the solution of our ME in a paradigmatic case. We choose the situation where the system-ancilla coupling is of the Jaynes-Cummings (JC) form. This, indeed, routinely occurs in quantum optics and is often used as an illustrative instance of various proposed approaches to NM dynamics.

This paper is structured as follows. In Section \ref{review}, we briefly review the theory leading to the discussed ME for incoherent swap intra-bath collisions. In Section \ref{temp}, we show how to include temperature in our framework. In Section \ref{JC}, we consider a system-ancilla interaction having the form of a JC coupling and explicitly work out the dynamical map entailed by our ME. Finally, in Section \ref{conc}, we draw our conclusions.

\section{Review of the collision model for incoherent partial swap}\label{review}

Our model comprises a system $S$, initially in state $\rho_0$, and a bath consisting of a collection of ancillas labeled with $i\ug1,2,...$. The initial overall state is $\sigma_0\ug \rho_0 \rho_{B0}$, where $\rho_{B0}$ is the bath initial state given by
\begin{equation}
\rho_{B0}\ug \ket{{\bf 0}}_B\!\bra{{\bf 0}} \label{in-state}\,\,,
\end{equation}
where $\ket{{\bf 0}}_B\ug\ket{{0}}_1\!\ket{{0}}_2\!\cdot\!\cdot\cdot$.
As mentioned, in our model inter-ancillary (AA) collisions are interspersed with system-ancilla (SA) ones. In the beginning, $S$ collides with {ancilla~1}. In standard (Markovian) CMs, $S$-2 collision would then follow, then $S$-3 and so on. This way, each ancilla would still be in the initial state $\ket{0}\!\bra{0}$ before colliding with $S$, thus fully ``unaware" of previous collisions. In contrast, in our model we assume that an extra AA collision between 1 and 2 occurs after $S$-1 but before $S$-2. Thereby, prior to its interaction with $S$, ancilla 2 will now be in a perturbed state in which information over past history of $S$ is imprinted. The process proceeds by mere iteration: once $S$-2 collision is over, a 2-3 interaction follows, then $S$-3, 3-4 etc. 

Each collision, either SA or AA, is described by a CPT quantum map affecting the degrees of freedom of the two involved particles. 
 Specifically,  the map for an SA collision involving the $i$th ancilla is defined as {the unitary coupling}  
 $\mathcal{U}_{Si}\sigma\ug e^{-i \hat{H}_{Si} t_c}\sigma {e^{i \hat{H}_{Si} t_c}}$, where $t_c$ and $\hat H_{Si}$ are respectively the collision time and interaction  Hamiltonian  (we set $\hbar\ug1$ throughout). 
Instead,  the  AA collision involving  the  ancilla $i$ to the $j$th one 
is  defined in terms of the CPT map 
\begin{eqnarray}
\mathcal{S}_{ij}\sigma&\ug& (1-p_s)\sigma\piu p_s \hat{S}_{ij}\sigma\hat{S}_{ij}\,\,\label{ipswap}
\end{eqnarray}
where $p_s$ is the swap probability while $\hat{S}_{ij}$ is the well-known swap operator \cite{NC} exchanging the states of $i$ and $j$ as $\hat{S}_{ij}\ket{\varphi}_i\!\ket{\psi}_j\ug \ket{\psi}_i\!\ket{\varphi}_j$. The steps of our discrete process are defined in a such way that the $n$th step terminates when both $(n$-1)--$n$ and $S$--$n$ collisions are over (but step $n\ug1$ ending after $S$--1 collision).
Hence, at the $n$th step the overall state is given by $\sigma_{n}=\left( \mathcal{U}_{Sn}\!\circ\!\mathcal{S}_{n,n\meno1}\!\circ \ldots \circ\!\mathcal{U}_{S2}\!\circ\!\mathcal{S}_{2,1}\!\circ\!\mathcal{U}_{S1}\right)\sigma_0$
(henceforth, {the} symbol ``$\circ$" {which describes super-operator composition} will be omitted). Let $\rho_n\ug{\rm Tr}_{B}(\sigma_n)$ be the $n$th -step state of $S$ (the partial trace is over all  the ancillas). Using the properties of the swap operator $\hat{S}_{i,i+1} {e^{-i \hat{H}_{Si} t_c}}\,\hat{S}_{i,i+1}\ug  {e^{-i \hat{H}_{S,i\piu1} t_c}}$ and $\hat{S}_{ij}\ket{\bf 0}_B\!\equiv\!\ket{\bf 0}_B$, one can easily show \cite{lavoro} that $\rho_n$ can be expanded in terms of all previous-step states of $S$ only. In the continuous limit, where due to $t_c\!\simeq\!0$ and $n\!\gg\!1$ the step number is replaced by the continuos time $t\ug nt_c$, such expansion for $\rho_n$ can be shown to give rise to a corresponding integro-differential ME. This reads
\begin{equation}\label{ME}
\frac{d\rho}{dt}=\Gamma\!\int_{0}^t \!\!dt'e^{-\Gamma t'} \mathcal{E}(t')\frac{\partial\rho(t\meno t')}{\partial  (t-t')}+e^{-\Gamma t}\frac{d \mathcal{E}(t) }{dt}\rho_0\,\,
\end{equation}
with the CPT map $\mathcal{E}(t)$ defined as
\begin{eqnarray}\label{map}
\mathcal{E}(t)\, \rho\ug\sum_{\nu}
{_n\langle \nu|
}\, {\rm e}^{-i\hat{H}_{Sn}t}\,|0\rangle_n\,\rho\,\left({_n\langle \nu|
}\, {\rm e}^{-i\hat{H}_{Sn}t}\,|0\rangle_n\right)^\dagger\,\,,
\end{eqnarray}
where each Kraus operator ${_n\langle \nu|
}\, {\rm e}^{-i\hat{H}_{Sn}t}\,|0\rangle_n$ is independent of $n$ since so is the form of $\hat{H}_{Sn}$ and the ancillas are all identical ($\{\ket{\nu}\}$ is a basis for the single-ancilla Hilbert space). Map $\mathcal{E}(t)$ is in general strongly NM since, evidently, it effectively describes the continuous coherent interaction of $S$ with a {\it single} ancilla. Indeed, our model interpolates between two extreme regimes depending on the value of $\Gamma$. When $\Gamma\ug0$, which can be shown to correspond to $p_s\ug1$ in the discrete model, swapping is perfect and as is immediate to see from \eq(\ref{ME}) the solution reads $\rho(t)\ug\mathcal{E}(t)\rho_0$. In such a case, then, \eq(\ref{map}) coincides with the process dynamical map $\Lambda(t)$ [this is defined through $\rho(t)\ug\Lambda(t)\rho_0$]. In the opposite limit $\Gamma\!\to\!\infty$, instead, it can be proven \cite{lavoro} that $\Lambda(t)\ug e^{\dot{\mathcal{E}}(0)t}$ with \eq(\ref{ME}) reducing to the Lindblad form $\dot{\rho}\ug \dot{\mathcal{E}}(0)\rho$, where $\dot{\mathcal{E}}(0)$ is a Lindlabian superoperator. Such limiting case indeed corresponds in the discrete picture to $p_s\ug0$, i.e., AA collisions do not occur at all [\cf\eq(\ref{ipswap})] and a standard fully Markovian CM is retrieved.

The general expression for $\Lambda(t)$ can be worked out as follows. Clearly, $\Lambda(t)$ obeys \eq(\ref{ME}) under the formal replacement $\rho\!\rightarrow\!\Lambda$. By taking the Laplace transform (LT) of such equation, this is easily solved as \cite{lavoro} 
 \begin{eqnarray} \label{ffd}
 \tilde{{\Lambda}}(s)=    \frac{ \tilde{\mathcal E}(s+\Gamma)}{{\mathcal I} - \Gamma\;\tilde{\mathcal E}(s+\Gamma)}\,\,
\end{eqnarray} 
where $\tilde{\Lambda}(s)$ and $\tilde{\mathcal{E}}(s)$ are the LTs of $\Lambda(t)$ and $\mathcal{E}(t)$, respectively.
Expanding \eq(\ref{ffd}) in powers of $\Gamma$ gives $ \tilde{{\Lambda}}(s)\ug\sum_{k=1}^\infty\left[\tilde{\mathcal{E}}(s\!+\!\Gamma)\right]^k\!\Gamma^{k\!-\!1}$, whose inverse LT is
  \begin{eqnarray} \label{ffd3}
\Lambda(t) \ug\mathcal{L}^{-1}[\tilde{\Lambda}(s)](t)\,\ug  \sum_{k=1}^\infty\Gamma^{k\!-\!1} \; \mathcal{L}^{-1}\![\tilde{\mathcal{E}}^k(s\!+\!\Gamma)](t)\;.
\end{eqnarray} 
Basic properties of LT allow to immediately calculate the inverse LT on the right-hand side as 
  \begin{eqnarray} \label{conv}
\!\mathcal{L}^{\meno1}\![\tilde{\mathcal{E}}^k\!(\!s\!+\!\Gamma\!)] \ug e^{-\Gamma t}\int_0^t \!\!\!dt_1\!\int_0^{t_1}\!\!dt_2\!\cdot\!\cdot\!\cdot\!\int_0^{t_{k\meno2}} \!\!d t_{k\meno1} 
\; \mathcal{E}(t_{k\meno1}\!) \mathcal{E}(t_{k\meno2}\!\meno \!t_{k\meno1}\!)\cdot\!\cdot\!\cdot\!\mathcal{E}(t\!\meno\! t_1\!).\,\,\, \,
\label{ltn}
\end{eqnarray} 
We have thus expressed $\Lambda(t)$ as a weighted series of multiple auto-convolutions of the CPT map $\mathcal{E}(t)$. Being a composition of CPT maps, each convolution [the integrand in \eq(\ref{ltn})] is CPT itself. Also, it is multiplied by a positive coefficient [\cf\eqs(\ref{ffd3}) and (\ref{ltn})], which yields complete positivity of map $\Lambda(t)$. Moreover, it is easily checked \cite{lavoro} that regardless of $\Gamma$ and $\mathcal E(t)$ ${\rm Tr}\left[\Lambda(t)\rho_0\right]\ug1$. Map $\Lambda(t)$ is thus {\it always} CPT.

\section{Including temperature} \label{temp}

As mentioned in the Introduction, the natural way to account for temperature $T$ is to replace the bath initial state \eq(\ref{in-state}) with a product of single-ancilla thermal states as
\begin{equation}\label{in-state-2}
\rho_{B0}\ug \bigotimes_i \rho_{i0}\ug\bigotimes_i\frac{{\rm e}^{-\beta\hat{H}_i}}{{\rm Tr}\left({\rm e}^{-\beta\hat{H}_i}\right)}\,\,,
\end{equation}
where $\beta\ug 1/(K T)$  ($K$ is the Boltzmann constant) and $\hat{H}_i$ is the single-ancilla free Hamiltonian {(whose operator form is independent of $i$)}. For $d$-dimensional ancillas, let  $\{\ket{\kappa}_i\}$ ($\kappa\ug0,...,d\meno1$) be the orthonormal set of eigenstates of $\hat{H}_i$ such that $\hat{H}_i\ket{\kappa}_i\ug\varepsilon_\kappa\ket{\kappa}_i$.  Then, the $i$th-ancilla thermal state $\rho_{i0}$ is in general a mixture of $\{\ket{\kappa}_i\!\!\langle \kappa|\}$ according to
\begin{equation}\label{rhoi0}
\rho_{i0}\ug\sum_{k=0}^{d\meno1}\frac{ {\rm e}^{-\beta \varepsilon_\kappa}}{Z}\ket{\kappa}_i\!\!\langle \kappa|\,\,,
\end{equation}
where 
$Z\ug\sum_{\kappa=0}^{d\meno1} {\rm e}^{-\beta \varepsilon_\kappa}$
is the partition function.
Clearly, for $\beta\!\to\!\infty$, namely zero temperature, state \eq(\ref{in-state-2}) reduces to a product of pure states such as \eq(\ref{in-state}). For non-zero temperatures, the ancillary initial state can be ``purified" \cite{NC} as follows. 
In general, given a single-particle mixture describing the state of a particle $A$, one can define a further fictitious particle $B$ in a way that $A$-$B$ are in a pure global state $\ket{\psi}_{AB}$ and $\rho_A\ug{\rm Tr}_B(\ket{\psi}_{AB}\!\bra{\psi})$. Following this scheme, we thus replace in our model each $d$-dimensional ancilla with a pair of identical $d$-dimensional ancillas such that only one of the two can interact with $S$ ($d$ is an arbitrary integer). We now use indexes $i_1$ and $i_2$ to label the ancillas really interacting with $S$ and the corresponding fictitious ones, respectively. Each $i_1$th ancilla is initially in state $\rho_{i_10}$ as given by \eq(\ref{rhoi0}). Correspondingly, the joint system comprising $i_1$ and the associated auxiliary ancilla $i_2$ is taken to be initially in the pure state
\begin{equation}\label{psi0}
\ket{\psi_0}_{{i_1i_2}}\ug{\sum_{\kappa=0}^{d\meno1} \sqrt{\frac{{\rm e}^{-\beta \varepsilon_\kappa}}{{Z}}}\ket{\kappa}_{i_1}\!\!\ket{\kappa}_{i_2}}\,\,,
\end{equation}
which evidently fulfills $\rho_{i_10}\ug{\rm Tr}_{i_2}(\ket{\psi_0}\!\!\bra{\psi_0})$. 

In the derivation of \eq(\ref{ME}) no assumption on either the dimensionality of the ancillas or the features of $\hat{H}_{Si}$ is necessary. Therefore, map $\mathcal{E}(t)$ now reads
\begin{eqnarray}\label{map2}
\mathcal{E}(t)\rho\ug\sum_{\nu_1\nu_2}
{\langle \nu_1\nu_2|
}\, e^{-i \hat{H}_{Si_1} t}\,|\psi_0\rangle\,\rho\left({\langle \nu_1\nu_2|
}\, e^{-i \hat{H}_{Si_1} t}\,|\psi_0\rangle\right)^\dagger\,\,,
\end{eqnarray}
where $\{|\nu_1\rangle\}_{i_1}$ ($\{|\nu_2\rangle\}_{i_2}$) is a basis for $i_1$ ($i_2$).
Using that $\hat{H}_{Si_1}$ does not affect $i_2$, we end up with 
\begin{eqnarray}\label{map3}
\mathcal{E}(t)\rho\ug\sum_{\kappa}{\frac{{\rm e}^{-\beta \varepsilon_\kappa}}{{Z}}}\left\{\sum_{\nu_1}
{\langle \nu_1|
}\, e^{-i \hat{H}_{Si_1} t}\,|\kappa\rangle\,\rho\left({\langle \nu_1|
}\, e^{-i \hat{H}_{Si_1} t}\,|\kappa\rangle\right)^\dagger\right\}\,\,,
\end{eqnarray}
where our notation emphasizes that the resulting map is a convex combination of maps defined in terms of initial pure single-ancilla states $\{|\kappa\rangle\}$. 
 We conclude that ME \eq(\ref{ME}) holds at finite temperature as well provided that map $\mathcal{E}(t)$ in \eq(\ref{map3}) is considered. Note that {\eq(\ref{map3}) also shows that $\mathcal{E}(t)$ can now be regarded as the map describing the reduced unitary dynamics associated with the time-evolution operator $e^{-i \hat{H}_{Si_1} t}$ and the initial $S$--$i_1$ state given by the tensor product of $\rho$ with \eq(\ref{rhoi0}). Hence, it is still true that for $\Gamma\ug0$ $S$ effectively behaves as it is continuously interacting with a single ancilla.}

\section{Dynamical map for Jaynes-Cummings coupling}\label{JC}

We assume that $S$ is a qubit \cite{NC}, i.e., a two-level system, and the SA coupling has the form of an $XY$ {\it isotropic interaction} corresponding to a JC model when $S$ and $A$ are on resonance \cite{JC}.
Each ancilla is modeled as a bosonic mode initially in the vacuum state, but because of conservation of the total number of excitations it behaves as an effective qubit as well. Let then $\{\ket{0}_
{S(i)}$,$\ket{1}_{S(i)}\}$ be a basis for $S$ (the $i$th ancilla). The SA Hamiltonian reads $\hat{H}_{Si}\ug\Omega \left(\hat{\sigma}_+ \hat{S}_{i-}\piu{\rm H.c.}\right)$, where $\Omega$ is a coupling rate while $\hat{\sigma}_{+}\ug\hat{\sigma}_-^\dag\ug\ket{1}_S\!\!\bra{0}$ and $\hat{S}_{i+}\ug \hat{S}_{i-}^\dag\ug\ket{1}_i\!\bra{0}$ are the usual spin-1/2 ladder operators. 
The most general initial state of $S$ reads $\rho_0\ug(1\meno p )\ket{0}_S\!\bra{0}\piu p \ket{1}_S\!\bra{1}\piu (r\ket{0}_S\!\bra{1}\piu{\rm H.c.})$, where $0\!\le\!\!p\!\!\le1$ and $\vert r\vert^2\!\le\!p(1\meno p)$. An amplitude damping channel (ADC) \cite{NC} transforms $\rho_0$ into
\begin{equation}\label{ejc}
\!\!\!\!\!\!\!\!\!\!\!\!\mathcal{A}(\eta)\rho_0\ug [1- \eta^2 p]\ket{0}_S\!\bra{0}+ \eta^2 p \ket{1}_S\!\bra{1}+ [ \eta r\ket{0}_S\!\bra{1}+{\rm H.c.}]\,\,,\,\,\,\,\,
\end{equation}
where $0\!\le\!\eta\!\le\!1$. Through a standard calculation, when the ancillas are initially all in $|0\rangle$ the map in \eq(\ref{map}) is given by $\mathcal{E}(t)\ug\mathcal{A}[\cos(\Omega t)]$. ADCs fulfill the composition property $\mathcal{A}(\eta_1)\mathcal{A}(\eta_2)\!\ug\!\mathcal{A}(\eta_1\eta_2)$. Using this in combination with \eqs(\ref{ffd3}) and (\ref{conv}) and introducing the rescaled time $\tau\ug\Omega t$, the dynamical map can be expressed in matrix form as
\begin{equation}\label{lambdajc}
\Lambda(\tau)\rho_0\ug \left( \begin{array}{cccc}
1-\beta_2(\tau)p&                     \beta_1(\tau)   r          \\
            \beta_1(\tau)r^*          &\beta_2(\tau) p \end{array} \right)
\end{equation}
with
\begin{eqnarray}\label{betat}
\!\!\!\!\!\!\!\!\!\!\!\!\!\!\!\!\!\!\!\!\!\!\!\!\!\!\!\!\!\!\!\!\!\!\!\!\beta_\ell(\tau)\ug e^{-\bar{\Gamma} \tau}\sum_{k=1}^\infty\bar{\Gamma}^{k-1}\!\!\!\int_0^\tau d\tau_1\!\!\int_0^{\tau_1}\!\!d\tau_2\cdot\!\cdot\!\cdot\!\!\int_0^{\tau_{k\meno2}} d \tau_{k\meno1} 
\cos(\tau_{k\meno1})^\ell \cos(\tau_{k\meno2}\meno \tau_{k\meno1})^\ell\cdot\!\cdot\!\cdot\!\cos(\tau\!\meno\! \tau_1\!)^\ell,\,\,
\end{eqnarray}
where $\bar{\Gamma}\ug\Gamma/\Omega$. Hence, functions $\beta_1(\tau)$ and $\beta_2(\tau)$ fully specify the dynamical map. We define the LT of a cosine power as $\tilde{c}_\ell(s)\ug\mathcal{L}[\cos(\tau)^\ell]$. Given that \eq(\ref{betat}) is formally analogous to \eq(\ref{ffd3}) with due replacements, in line with \eq(\ref{ffd}) the LT of $\beta_\ell(\tau)$ is evidently given by
\begin{eqnarray}\label{betas}
\tilde{\beta}_\ell(s)\ug\mathcal{L}[\beta_\ell(\tau)]\ug \frac{\tilde{c}_\ell(s\piu\bar{\Gamma} )}{1-\bar{\Gamma} \tilde{c}_\ell(s\piu\bar{\Gamma} ) }\,\,.
\end{eqnarray}
Now, by using $\tilde{c}_1(s)\ug s/(s^2\piu1)$ along with $\tilde{c}_2(s)\ug (s^2\piu2)/[s(s^2\piu4)]$ and anti-transforming we find 
\begin{eqnarray}
\beta_1(\tau)\ug e^{-\frac{\bar{\Gamma} \tau}{2}} \left[\frac{\bar{\Gamma} \sinh
   \left(\frac{1}{2} \sqrt{\bar{\Gamma}^2\meno4}
   \,\tau\right)}{\sqrt{\bar{\Gamma}^2\meno4}}+\cosh
   \left(\frac{1}{2} \sqrt{\bar{\Gamma}^2\meno4}
 \,  \tau\right)\right]\,\,,\label{beta1}\\
\beta_2(\tau)\ug \sum_{i=1}^3 A_i {\rm e}^{i\alpha_i\tau}\label{beta2}\,\,,
\end{eqnarray}
where
\begin{eqnarray}\label{coeffs}
\alpha_1\ug\frac{\left(\bar{\Gamma}-\sqrt[3]{\bar{\Gamma}^3+3\delta+9 \bar{\Gamma}}\right)^2-12}{3
   \sqrt[3]{\bar{\Gamma}^3+3 \delta+9
   \bar{\Gamma}}}\,\,,\nonumber\\
   \alpha_2\ug   \alpha_3^*\ug\frac{1}{6} \left(i \left(\sqrt{3}+i\right)
   \sqrt[3]{\bar{\Gamma}^3+3 \delta+9
   \bar{\Gamma}}-\frac{\left(1+i \sqrt{3}\right)
   \left(\bar{\Gamma}^2-12\right)}{\sqrt[3]{\bar{\Gamma}^3+3
   \delta+9 \bar{\Gamma}}}-4 \bar{\Gamma}\right)\,\,,\,\,\,\,\,\,\,\,\,\,\,\,\,\,\,\,\nonumber\\
   A_1\ug\frac{2 \bar{\Gamma}
   \alpha_1+\alpha_1^2+\bar{\Gamma}^2+2}{\alpha_1^2+|\alpha_2|^2-2
   \alpha_1
   {\rm Re}(\alpha_2)}\,\,,\nonumber\\
   A_2\ug  A_3^*\ug\frac{i \left(2 \bar{\Gamma}
   \alpha_2+\alpha_2^2+\bar{\Gamma}^2+2\right)
   }{2 (\alpha_1\meno\alpha_2)
   {\rm Im}(\alpha_2)}\,\,\nonumber
\end{eqnarray}
with $\delta\ug  \sqrt{6  \bar{\Gamma}^4-39  \bar{\Gamma}^2+192}$.
We have checked that $\beta_1(\tau)$ and $\beta_2(\tau)$, as given by \eqs(\ref{beta1}) and (\ref{beta2}), fulfill the conditions $0\!\le\!\beta_2(\tau)\!\le\!1$ and $\beta_1(\tau)^2\!\le\!\beta_2(\tau)$ regardless of $\Gamma$ and $\tau$. This confirms that map \eq(\ref{lambdajc}) is unconditionally CPT in accordance with our general theory.

\section{Conclusions}\label{conc}

In conclusion, we have briefly reviewed a CM-based framework to tackle non-Markovian dynamics. The model differs from standard Markovian CMs in that extra inter-ancillary collisions are added between next system-ancilla interactions. Here, we have focussed on the case where each ancilla-ancilla collision is described by an incoherent partial swap. The corresponding dynamics is exactly described by a unconditionally CPT ME. We have complemented the related theory with two further developments. First, we have shown that the case where the bath is initially in a thermal state, instead of a pure one,  is in fact already encompassed in the theory provided that one suitably redefines each ancilla as a bipartite system (initially entangled). Next, we have illustrated for the first time a paradigmatic instance where the solution of the ME can be calculated explicitly in a compact form. We have chosen the case where the system-ancilla interaction is described by a JC coupling.

\section*{Acknowledgments}

We acknowledge useful discussions with G. M. Palma and support from the MIUR through the FIRB-IDEAS project RBID08B3FM.

\end{document}